\documentstyle[aps,pre,multicol]{revtex}

\def\beginwide{
        \end{multicols} \vspace*{-0.5cm} \noindent
        \rule{3.5in}{.1mm}\rule{.1mm}{5mm} \widetext \medskip }
\def\beginwidetop{
        \end{multicols} \vspace*{-0.5cm} \noindent
        \widetext \medskip }
\def\endwide{
        \hspace*{3.5in}~\rule[-5mm]{.1mm}{5mm}\rule{3.5in}{.1mm}
        \begin{multicols}{2} \vspace*{-1.0cm} \noindent }
\def\endwidebottom{
        \begin{multicols}{2} \vspace*{-1.0cm} \noindent }

\tighten
\draft

\begin{document}

\title{Constant rate linear interface depinning and self-organized 
criticality}
\author{Alexei V\'azquez and Oscar Sotolongo-Costa}
\address{Department of Theoretical Physics, Faculty of
Physics, Havana University, Havana 10400, Cuba}

\maketitle

\begin{abstract}

The precise determination of the universality classes in self-organized
critical phenomena (SOC) is still an unsolved problem. Different SOC models
like sandpile, linear interface depinning, and the Barkhausen effect have
been investigated independently. In the present work we demonstrate that
these models can all be mapped into a linear interface depinning model driven
at constant rate. The model is found to belong to the universality class of
constant force linear interface depinning above the depinning transition,
with an upper critical dimension $d_c=4$. Results are compared with numerical
simulations, experiments, and previous theoretical works reported in the
literature. In this way we demonstrate a precise connection between different
SOC models which display the same universal behavior.

\end{abstract}


\begin{multicols}{2}
\narrowtext

\section{Introduction}
\label{sec:intro}

About one decade ago Bak, Tang and Wiesenfeld (BTW) introduced the idea of
self-organized criticality (SOC) \cite{bak} to describe the critical behavior
of a vast class of driven dissipative systems. The dynamics of SOC systems is
characterized by long periods of quasi-equilibrium alternating with sudden
rearrangements (avalanches) which may expand all the system. Avalanche
dynamics has been observed in granular sandpiles \cite{bak,sandpiles}, the
Barkhausen effect \cite{alessandro,urbach,durin1,cizeau,durin2},
superconducting vortex piles \cite{superconductors}, earthquakes
\cite{earthquakes}, etc.

As usual the mean-field (MF) theory gave the first insight into SOC phenomena
\cite{tang,zapperi,vespignani1}. We also count with some exact results for 
Abelian sandpile models \cite{dhar,priezzhev}, real space \cite{pietronero}
and momentum space \cite{diaz-guilera} renormalization group (RG)
calculations, and singular diffusion equations \cite{carlson2}. More recently
the analogy with other non-equilibrium systems like systems with many
absorbing states \cite{dickman,vespignani2}, directed percolation
\cite{tadic,maslov,vazquez1}, and linear interface depinning (LID) \cite{hwa,paczuski1,vazquez2,lauritsen} has been exploited. This accumulated experience allowed a more precise formulation of SOC phenomena. It is now well known that the absence of contro
l parameters, as it was believed at the early state, is not completely true. Some hidden parameters like the driving rate should be fine-tuned in order to reach the critical state.

However the precise determination of the universality classes in SOC
phenomena is still an unsolved problem. For instance, it is not clear if
deterministic and stochastic sandpile models belong to the same universality
class. Some numerical simulations \cite{lubeck,chessa}, real space RG
calculations \cite{pietronero} and field theory \cite{vespignani2} pointed to
the affirmative answer but other numerical simulations \cite{biham} disagree.
The problem becomes more difficult since in deterministic models multifractal
scaling, instead of finite size scaling, is satisfied \cite{stella}. On the
other hand, some authors have pointed out the existence of universal behavior
between sandpile and LID models \cite{paczuski1,lauritsen}. Paczuski and
Boettcher \cite{paczuski1} mapped the random slope threshold sandpile model
\cite{random-slope} into a model of LID where the interface is pulled at one end. More recently we have extended this analogy to critical height models \cite{vazquez2}, which are mapped into a LID model driven at constant rate. In this case we have obtain
ed the complete set of scaling exponents using previous RG analysis for constant force LID models. Lauritsen and Alava \cite{lauritsen} have also considered the analogy between critical height models and LID models. However in their work the way in which 
driving and dissipation take place are not well specified. The analysis is limited to a qualitative level and no rigorous determination of the scaling exponents was provided, thought they count with the previous work by us \cite{vazquez2}.

The way in which the interface is driven seems to be an important condition
to obtain an ordinary depinning transition or SOC behavior. The interface may
be driven either by extremal dynamics \cite{paczuski2,maslov2,roux}, by
constant force \cite{leschhorn,dong,makse0} or by constant rate
\cite{vazquez2}. While constant force models has been extensively studied in 
the literature extremal dynamics and constant rate models are less known.
However, in the last years, extremal dynamics and constant velocity models
have gained more attention due to its relation with the theory of
self-organized criticality (SOC). However, constant rate, and not constant
force or constant velocity, is the predominant driving mechanism in
experiments where avalanche dynamics is observed. Examples are found in
granular sandpiles where the number of grains added to the system increases
linearly with time \cite{bak}, in Barkhausen and superconducting avalanches
experiments where the applied magnetic field is increased at constant rate,
and in earthquake dynamics where some tectonic plate dynamics gives a
constant rate of stress increase leading to a stick-slip motion of other
plates.

In the present work we investigate the existence of universality in different
SOC models proposed in the literature, which can be mapped into a LID model
driven at constant rate. The LID model driven at constant rate is taken here
as the prototype of this class of SOC models because it allows some
analytical treatment, which is very difficult for instance in sandpile
models. The existence of a different driving mechanism does not change the
universality class; we show that the constant rate model belongs to the
universality class of constant force LID model, with an upper critical
dimension $d_c=4$. Contrary to previous works, our analysis is not limited to
a qualitative level but it is supported by a complete mapping of different
SOC models into a LID model. The complete set of scaling exponents, including
the avalanche exponents, are computed using previous RG analysis developed
for constant force LID models. Part of these results are already contained in
\cite{vazquez2}.

The paper is organized as follows. In section \ref{sec:model} we introduce
the LID model driven at constant force and review some of the results
obtained from RG analysis. The LID model driven at constant force has been
already investigated in the literature, we just present some important
aspects which are determinant for the analysis developed in the next section.
In section \ref{sec:constant-rate} we consider the constant rate LID model
and illustrate how different SOC models can be mapped into this model. The
phase diagram of this model is investigated in section \ref{sec:phase}. The
comparison of our results with experiments, numerical simulations and other
theoretical approaches proposed in the literature is discussed in section
\ref{sec:disc}. Finally the summary and conclusions are given in section 
\ref{sec:summ}.

\section{Constant force LID models}
\label{sec:model}

There are many physical situations where one is interested in an interface
dynamics. Examples are found in the domain wall dynamics
\cite{alessandro,cizeau}, the displacement of one fluid by another inside a 
porous media \cite{rubio,martys1}, contact line depinning \cite{ertas}, and
more \cite{halpin}. In general a $d$-dimensional self-affine interface,
described by a single-valued function $h(x,t)$, evolves in a
$(d+1)$-dimensional medium. Usually some kind of disorder affects the motion
of the interface leading to its roughening.  Earlier studies \cite{kardar}
focuse on time-independent uncorrelated disorder but most recent studies
analyze the motion of interfaces under quenched disorder
\cite{martys1,martys2,ji}. In the presence of quenched disorder and constant 
force driving two universality classes has been found \cite{amaral}. One is
described by the Kardar-Parisi-Zhang equation \cite{kardar} with quenched
noise. In this case the interface is pinned by paths on a directed
percolation cluster of pinning sites \cite{tang1}. The second class is
described by the Edwards-Wilkinson equation \cite{edwards} with quenched
noise, usually known as LID models.

In LID models the interface height satisfies the equation of motion
\begin{equation} 
\lambda \partial_t h=\Gamma\nabla^2h+F+\eta(x,h), 
\label{eq:a1} 
\end{equation}
where $\partial_t$ denotes the partial time derivative and $\nabla$ is a
$d$-dimensional Laplacian. $\lambda$ is a viscosity coefficient, $\Gamma$ the
surface tension and $F$ is a constant force acting on the interface.
$\eta(x,h)$ is a random pinning force associated with the existence of random
pinning centers in the $(d+1)$-dimensional environment. In general it is
assumed that $\eta(x,h)$ is a Gaussian noise uncorrelated in the space $x$,
with zero mean and noise correlator
\begin{equation}
\langle\eta(x,h)\eta(x^{\prime},h^{\prime})\rangle =
\delta^d(x-x^{\prime})\Delta(h-h^{\prime}). 
\label{eq:a2} 
\end{equation}
where $\Delta(-h)=\Delta(h)$ is a symmetric function, with a fast decay to
zero beyond the characteristic length of the pinning centers.

The existence of pinning centers will carry as a consequence that the
interface will be not smooth. The roughness of the interface is characterized
by the height-height correlation function
\begin{equation}
\langle[y(x,t)-y(0,0)]^2\rangle \sim
|x|^{2\zeta}g(t/|x|^z),
\label{eq:c7a}
\end{equation}
where $\zeta$ and $z$ are the roughness and dynamic scaling exponents,
respectively, and $g(x)$ a scaling function with the asymptotic behaviors
$g(x)\sim1$ for $x\ll1$ and $g(x)\sim x^{2\zeta/z}$ for $x\gg1$. The scaling
exponents $z$ and $\zeta$ are related with the scaling behavior of the
interface fluctuations just at the critical state and therefore we expecct
that they are independent on the way the system is driven out from the
critical state.

A depinning transition takes place at certain critical force $F_c$ determined
by the disorder. For $F<F_c$ the interface is pinned after certain finite
time while for $F>F_c$ it moves with finite average velocity $v$. Above $F_c$
one looks for a solution of eq. (\ref{eq:a1}) as an expansion around the flat
co-moving interface $vt$, i.e. $h(x,t)=vt+y(x,t)$, resulting the equation of
motion for $y(x,t)$
\begin{equation} 
\lambda \partial_t y=\Gamma\nabla^2y+F-\lambda v+\eta(x,vt+y). 
\label{eq:1} 
\end{equation}
$v$ is obtained self-consistently using the constraint $\langle
y(x,t)\rangle=0$. The noise term in eq. (\ref{eq:1}) makes this equation
non-linear requiring a RG analysis to determine the scaling behavior in the
neighborhood of the depinning transition. This has been done using different
approaches by Natterman {\em et al} \cite{nattermann} and Narayan and Fisher
\cite{narayan2,narayan}. The order parameter of the depinning transition is the average interface velocity. For $F<F_c$ we have $v=0$ while for $F>F_c$ the interface moves with the average interface velocity $v\sim(F-F_c)^\beta$, with
\begin{equation}
\beta=\nu(z-\zeta),
\label{eq:2b}
\end{equation}
where $\nu$ is the correlation length exponent. 

The upper critical dimension is $d_c=4$ and the scaling exponents can be
obtained through a functional RG analysis. Below the upper critical dimension
it results that
\begin{equation}
\zeta=\frac{4-d}{3},\ \ \ \ z=2-\frac{2}{9}(4-d).
\label{eq:3}
\end{equation}
Moreover, near the critical state the static susceptibility and crorrelation
length diverges according to $\chi\sim(F-F_c)^{-\gamma}$ and
$\xi\sim(F-F_c)^{-\nu}$, respectively, where the scaling exponents $\gamma$
and $\nu$ satisfy the scaling relation
\begin{equation}
\frac{\gamma}{\nu}=2.
\label{eq:2a}
\end{equation}

Alternative to the continuum equation (\ref{eq:a1}) one may consider the
following discrete model \cite{leschhorn}. In a $d+1$ hypercubic discrete
lattice each site is labeled by index $i$ and height $h_i$. Discrete time
steps $t=0,1,\ldots$ are taken. The force acting on each site and the
evolution of the interface height $h_i(t)$ are given by
\begin{eqnarray}
F_i=\sum_{nn}h_j-2dh_i+F+\eta_i(h_i),
\nonumber\\
h_i(t+1)-h_i(t)=\Theta(F_i),
\label{eq:a3}
\end{eqnarray}
where $\Theta(x)$ is the Heaviside unit step function, $nn$ denotes that the
sum runs over the $2d$ nearest neighbors and $\sum_{nn}h_j-2dh_i$ is a
discretized Laplacian. On each step $t$ and at every site where $F_i>0$
($\Theta(F_i)=1$) the interface is advanced in parallel by one unit. Thus,
only at sites where $F_i>0$ the interface is active. The average interface
velocity is thus given by (in non-dimensional variables)
\begin{equation}
v=\rho_a,
\label{eq:a4}
\end{equation}
where $\rho_a$ is the density of active sites. In the continuous
representation this expression is equivalent to
\begin{equation}
\langle \partial_t h(x,t)\rangle=\langle \rho_a(x,t)\rangle,
\label{eq:a4a}
\end{equation}
where $\rho_a(x,t)$ is the coarse-grained density of active sites.

LID models are Abelian, in the sense that the order in which sites advance is
not important \cite{paczuski2}. If a site $i$ is active then it will transfer
energy to its nearest neighbors, which at the same time may become active,
and so on, an avalanche is generated.  It is thus possible that at certain
time step $t$ there will be more that one active site. These sites will be
updated in parallel according to the evolution rules described above.
However, the process of toppling can never transform any active site,
different from itself, in inactive and, therefore, the other active sites
will remain active. On the other hand, the energy transferred to its
neighbors is constant, independent of the total force at this site. Hence,
the order in which these sites are updated is not important.

\section{Constant rate LID model}
\label{sec:constant-rate}

In most systems which are expected to exhibit SOC the external field, instead
of being constant, increases linearly with time. For instance, in sandpiles
grains are usually added at constant rate so that the total number of grains
added to the pile up to time $t$ increases linearly with $t$. A more evident
example is found in magnetic noise measurements in ferromagnetic and
superconductor materials where the applied magnetic field increases linearly
with time. Hence, if we are trying to map any of these systems into a LID
model we can not assume a constant force, which will not be in correspondence
with the picture described above.

Motivated by this fact we propose a LID model where the force increases
linearly with time. However, in order to reach an stationary state we most
include a restoring force which balance the external field. If the force
increases linearly with time ($F=ct$) then after certain time it will
overcome the critical force $F_c$ and the interface will start moving with a
velocity $v(F=ct)$, which is time dependent. Hence the system could not reach
a constant velocity stationary state. This problem can be solved adding a
restoring force which balance the external driving and leads to an stationary
state. This can be done in different ways. For instance, one may consider a
local restoring force linear in $h$ resulting
\begin{equation}
F=ct-\epsilon h,
\label{eq:a2d}
\end{equation}
In this case the interface is driven by an external force which increases at
rate $c$ and a restoring force which pull the interface to the substrate. The
coefficient $\epsilon$ is a measure of the strength of the restoring force.
Now, suppose that at the initial state $h(x,0)=0$. With increasing time the
external force $F(t)=ct$ will increase. At certain time $F>F_c$ and the
interface will start moving with an average velocity $v$, i.e. $\langle
h\rangle=vt$. On the other hand, just when $h$ is finite the restoring force
will start pulling the interface to the substrate $h=0$. On average the
restoring force will be $F_R(t)=-\epsilon vt$. An stationary state will be
obtained when the restoring and pinning forces balance the external driving
force ($ct=F_c+\epsilon vt$) resulting, in the stationary state, the average
interface velocity $v=c/\epsilon$. We can also have a model with $\epsilon=0$
but with the interface pinned at the boundary, i.e. $h=0$ at the boundary. In
this case it is expected that the interface develop a parabolic profile so
that the surface tension balance the action of the driving force.

Alternative, instead of the local restoring force in eq. (\ref{eq:a2d}) one
may consider a golbal restoring force as follows
\begin{equation} 
F=ct-\epsilon\int\frac{d^dx^\prime}{L^d}h(x^\prime,t),
\label{eq:a5} 
\end{equation}
The qualitative analysis developed below will be also valid for this case.
Moreover, we have recently shown that both restoring forces lead to the same
critical behavior \cite{vazquez3}. In the following we will only consider the
local restoring force.

Next we proceed to show how different SOC systems can be mapped into a
constant rate LID model.

\subsection{The Barkhausen effect}

An inmediate realization of the equation of motion (\ref{eq:a1}) with a local
force given either by eq. (\ref{eq:a2d}) or (\ref{eq:a5}) is the dynamics o
domain walls. It is well known that domain walls in ferromagnets move
following an irregular motion in response to changes in an applied external
magnetic field, leading to discrete jumps in the magnetization, a phenomenon
known as the Barkhausen effect. In fact Urbach {\em et al} \cite{urbach}
considered a model where the domain wall dynamics is described by a LID
model, with either a local o global restoring force. More recently Cizeau
{\em et al} \cite{cizeau} have shown that the model by Urbach {\em et al} is
only valid if long-range dipolar interactions are neglected, an approximation
which is valid in soft magnetic materials \cite{durin2}. Hence, the dynamics
of a domain wall in a soft magnetic material can be described by a LID model
driven at constant rate.

\subsection{Sandpile models}

In cellular automaton sandpile models a discrete or continuous variable
$z_i$, height or energy, is defined in a $d$-dimensional lattice. The
dynamical evolution of $z_i$ is defined by two evolution rules: adding and
toppling. Different adding and toppling rules may be defined leading to
different models \cite{bak,sandpiles}. In particular we consider the
following rules
\begin{itemize}
\item adding: on each step each site receives a grain from the driving field 
with probability $c$;
\item toppling: if at certain site $z_i>z_c=2d-1$ then $z_i\rightarrow 
z_i-2d-\epsilon+\eta_i$ and $z_{nn}\rightarrow z_{nn}+1$ at the $2d$ nearest 
neighbors;
\end{itemize}
$\epsilon$ is the average dissipation rate per toppling event and $\eta_i$ is
a noise, which may have different origins. If $\epsilon>0$ the toppling rule
is non-conservative and the model have bulk dissipation. On the contrary, if
$\epsilon=0$ the toppling rule is conservative and one has to assume open
boundary conditions to balance the input of grains from the external field.
For instance, Chessa {\em et al} \cite{chessa} considered a nonconservative
sandpile model where the toppling site loses its energy with probability $p$
without transferring it to its neighbors. This corresponds to an average
dissipation rate per toppling event $\epsilon=2dp$, while $\eta_i$ will
reflect the stochastic nature of the local dissipation. Another example is
found in random threshold models \cite{christensen}. In this case after each
toppling event a new random threshold is assigned (a new critical slope in
the model by Christensen {\em et al} \cite{christensen}), which is equivalent
to introduce the noise $\eta_i$. On the other hand, at a coarse-grained level
the noise $\eta_i$ will appear due to elimination of microscopic degrees of
freedom. Hence, at this level, it will be present no matter if the model is
stochastic or deterministic. In other words, we expect that deterministic and
stochastic models belongs to the same universality class.

In the original BTW model $\epsilon=0$ and dissipation takes place at the
boundaries, while grains are added only after all sites become stable, i.e.
$z_i<z_c$ at all sites. However, it has been shown
\cite{vespignani1,vespignani2,chessa,barrat} that the BTW model and the 
model with bulk dissipation lead to the same critical behavior provided
$\epsilon\sim L^{-2}$, where $L$ is the lattice size, and $c\rightarrow0$. In
the limit $c=0^+$ we have separation of time scales between avalanche
duration and energy addition, as assumed in the BTW model. Moreover, as
discussed above, the noise $\eta_i$ will appear at a coarse-grained level.

Paczuski and Boettcher \cite{paczuski1} noted that a particular critical
slope sandpile model can be mapped into a LID model where $h_i(t)$ is the
number of toppling events at site $i$ up to time $t$. Their analysis was
limited to one dimension but can be extended to larger dimensions and models
with critical height rules \cite{vazquez2,lauritsen}. If at $t=0$ we have
$z_i=0$ at all sites then, at time step $t$, $z_i$ and $h_i$ are related via
\begin{equation}
z_i=\sum_{nn}h_j-2dh_i+ct-\epsilon h_i+\eta_i(h_i),
\label{eq:a7}
\end{equation}
where $\sum_{nn}h_j$ gives the number of grains received from nearest
neighbors, $2dh_i$ the number of grains transferred to nearest neighbors,
$ct$ the number of grains received from the driving field, and $-\epsilon
h_i$ the number of dissipated grains. When a site topples $h_i\rightarrow
h_i+1$ and therefore the interface profile $h_i(t)$ will always advance in
the positive direction. Since a site topples only when $z_i>z_c$ then
\begin{equation}
h_i(t+1)-h_i(t)=\Theta(z_i-z_c),
\label{eq:a8a}
\end{equation}

Instead of follows the evolution of the variables $z_i$, which are equivalent
to the force in LID problems, one may follow the evolution of $h_i$, which is
equivalent to the interface height in LID models. Notice that the evolution
equations (\ref{eq:a7}) and (\ref{eq:a8a}) for sandpile models are equivalent
to the discrete variant of LID models in eq. \ref{eq:a3}), taking
$F_i=z_i-z_c$. Hence, after coarse-graining, we obtain an equation of motion
for $h(x,t)$ like (\ref{eq:a1}) with $F$ given by eq. (\ref{eq:a2d}).
Actually there will be an additional term $-z_c$ in the right hand side of
eq. (\ref{eq:a1}) but it does not carry any changes in the critical behavior
so that one can work without considering this term.

\section{Phase diagram and scaling exponents}
\label{sec:phase}

We have shown that different SOC models can be mapped into a LID model where
the force increases at constant rate and a restoring force pulls the
interface to the substrate. In this section we solve this model using
previous results for the constant force variant. More precisely, we
investigate the dynamics of an interface described by the equation of motion
(\ref{eq:a1}) with $F$ given by eq. (\ref{eq:a2d}), i.e.
\begin{equation} 
\lambda \partial_t h=\Gamma\nabla^2h+ct-\epsilon h+\eta(x,h). 
\label{eq:b0} 
\end{equation}
As we discussed above, in this case the interface will never be pinned but
moves with a finite average velocity $v$. It is thus expected that the
dynamics will be similar to that observed in constant force models above the
critical force. Based on this supposition we will perform a suitable change
of reference in order to obtain an equation similar to that for the
fluctuations around the average in constant force models, eq. (\ref{eq:1}).

\subsection{case $\epsilon=0$}

Let us first consider the case of boundary pinning which corresponds with
$\epsilon=0$ and $h=0$ at the boundary. We look for a solution in the form
\begin{equation}
h(x,t)=h_0(x,t)+y(x,t),
\label{eq:b1}
\end{equation}
so that when we substitute this expression in eq. (\ref{eq:b0}) the constant
rate force will be replaced by a constant force. This constraint will be
satisfied taking $h_0(x,t)$ as the solution of the problem
\begin{equation}
\nabla^2h_0+ct=F.
\label{eq:b2}
\end{equation}
with the boundary condition $h_0=0$. The solution of this equation can be
easily obtained assuming radial symmetry $h_0=h_0(r,t)$ with $0<r<R$, where
$R$ is the system radius. It results that
\begin{equation}
h_0(r,t)=\left(1-\frac{r^2}{R^2}\right)\frac{ct-F}{2\Gamma d}R^2.
\label{eq:b2a}
\end{equation}
We have also analyzed the case of free boundary conditions $dh_0(R,t)/dr=0$,
resulting that eq. (\ref{eq:b2}) have no solution. In this case the system
will not reach an stationary state. This result supports our previous
analysis about the need of a restoring force which balance the external
driving of the interface.

Substituting eq. (\ref{eq:b1}) in eq. (\ref{eq:b0}) and taking the limits
$r\ll R$ and $ct\gg F$ we obtain eq. (\ref{eq:1}) with
\begin{equation}
v\approx\frac{\partial h_0}{\partial t}\approx\frac{cR^2}{2\Gamma d}.
\label{eq:b4}
\end{equation}
Notice that assuming $r\ll R$ and $ct\gg F$ we have approximated $h$ by
$vt+y$ in the quenched noise $\eta(x,h)$. However, the information contained
in $h_0$ is not completely lost because we have the constant force $F$ in the
right hand side of eq. (\ref{eq:1}). These approximations will be valid for
very long times (stationary state) and far from the boundary. $v$ is then the
average interface velocity in the stationary state, which depends on system
size $R$. When increasing system size the bulk increases faster than the
surface. Hence, since the interface is only fixed at the boundary the
concentration of points where the interface is fixed will decrease with
increasing system size resulting on an increase of the interface velocity.

The constant force $F$ has not been specified. It will be obtained
self-consistently imposing the constraint $\langle y(x,t)\rangle=0$. In
constant force models $F$ is a fixed parameter while $v$ is obtained
self-consistently from the equation of motion. However in the present model
$v$ is given by eq. (\ref{eq:b4}) while $F$ is the undetermined parameter.
From the constant force variant it is known that a force $F$ ($F>F_c$) gives
an average interface velocity $v\sim(F-F_c)^\beta$. Hence, in the constant
rate model, to obtain a velocity $v$ we should have the force
\begin{equation}
F(v)=F_c+\text{const.}v^{1/\beta}.
\label{eq:b5}
\end{equation}
In this way, $F(v)$ is obtained self-consistently imposing that the interface
moves with an average velocity given by eq. (\ref{eq:b4}). In spite of this
difference both models have the same critical behavior.

To reach the critical force $F_c$ we must fine-tune $v$ to zero. For $v>0$
there is a characteristic length $\xi\sim v^{- \nu/\beta}$. Now, the average
interface velocity scales as $v\sim cR^2$ and therefore when we take the
thermodynamic limit $R\rightarrow\infty$ we must fine-tune $c$ to zero to
reach the critical state. If $\xi\gg R$ then the system size $R$ will be the
only characteristic length and the system will be in a critical state. This
condition is satisfied if $c={\cal O}(R^{-2+\nu/\beta})$ when
$R\rightarrow\infty$.

In the case of boundary pinning the average density of active sites, computed
from eqs. (\ref{eq:a4a}), (\ref{eq:b1}) and (\ref{eq:b2a}), is given by
\begin{equation}
\langle\rho_a(r,t)\rangle=\left(1-\frac{r^2}{R^2}\right)\frac{c}{2\Gamma d}
R^2.
\label{eq:b5a}
\end{equation}
This expression is identical, except for some constant factors, to that
obtained in the field theory by Vespignani {\em et al} \cite{vespignani2}. It
just reflects the balance between the driving force and the boundary pinning
(the driving field and boundary dissipation in sandpile models). As one can
see from eq. (\ref{eq:b5a}) the average density develops a parabolic profile,
which is a consequence of the pinning at the boundary. This parabolic profile
has been observed in recent numerical simulations by Barrat {\em et al}
\cite{barrat}.

\subsection{Case $\epsilon>0$}

Now we consider the case of a local restoring force. We again look for a
solution in the form of eq. (\ref{eq:b1}) which the same constraints, i.e. it
replaces the constant rate force by a constant force $F$. This constraint is
satisfied if $h_0(x,t)$ is the solution of the problem
\begin{equation}
\nabla^2h_0-\epsilon h_0+ct=F.
\label{eq:b6}
\end{equation}  
with the corresponding boundary conditions. Independently of the boundary
conditions assumed, the solution of this equation exists. If one assumes the
fixed boundary condition then one will obtain a parabolic profile as in the
$\epsilon=0$ case. On the contrary, we will consider periodic or free
boundary conditions. In this case we obtain
\begin{equation}
h_0(x,t)=\frac{ct-F}{\epsilon}.
\label{eq:b8}
\end{equation} 

Substituting eq. (\ref{eq:b8}) in eq. (\ref{eq:b0}) and taking the limit
$vt\gg F/\epsilon$ we obtain
\begin{equation} 
\lambda \partial_t y=\Gamma\nabla^2y-\epsilon y+F-\lambda v+\eta(x,vt+y). 
\label{eq:b10} 
\end{equation}
with
\begin{equation}
v=\frac{c}{\epsilon}.
\label{eq:b9}
\end{equation}
Again assuming $vt\gg h_0(x)$ we have approximated $h$ by $vt+y$ in the the
quenched noise $\eta(x,h)$. Eq. (\ref{eq:b10}) is quite similar to that
obtained for the fluctuations around the flat co-moving interface in constant
force LID models, eq. (\ref{eq:a1}). The only difference is found in the term
$-\epsilon h$, which accounts for the local restoring force with strength
$\epsilon$. However, in the limit $\epsilon\rightarrow0$ the system will show
the same critical behavior as in constant force models. The term $-\epsilon
y$ only affects the linear part of the bare propagator while non-linear
terms, which are responsible for loop-corrections, remain identical. The
critical exponents $z$ and $\zeta$ obtained from the RG analysis will thus be
the same as those obtained for the constant force case. However the phase
diagram will show a complex structure.

For $v>0$ and $\epsilon>0$ the system is driving out the critical state. Now
in addition to the characteristic length $\xi\sim v^{\nu/\beta}$ we have
another characteristic length associated with the restoring force
$\xi_\epsilon\sim\epsilon^{-\nu_\epsilon}$, where $\nu_\epsilon$ is
calculated below. The phase diagram $(v,\epsilon)$ will have different
regions depending on the ratio between these two characteristic lengths. We
then define the characteristic velocity $v_\epsilon$ as the velocity where
these two characteristic lengths become identical. Thus taking
$\xi\sim\xi_\epsilon$ we obtain that
$v_\epsilon\sim\epsilon^{\beta_\epsilon}$ with
\begin{equation}
\beta_\epsilon=\frac{\nu_\epsilon}{\nu}(z-\zeta).
\label{eq:b11}
\end{equation}

In the region $v>v_\epsilon$ ($\xi>\xi_\epsilon$) $\xi$ is the only
characteristic length, as in the case $\epsilon=0$. However, we cannot say
that the effect of the restoring force disappears because the average
interface velocity depends on $\epsilon$ (see eq. (\ref{eq:b9})). There will
be a complete equivalence if we take $\epsilon\sim R^{-2}$. In other words, a
model with bulk dissipation with $v>v_c$ and $\epsilon\sim R^{-2}$ is
equivalent to a model without bulk dissipation and open boundaries. This
equivalence was already pointed out by Vespignani {\em et al} using MF
\cite{vespignani1} and field \cite{vespignani2} theory approaches.

On the contrary, in the region $v<v_\epsilon$ ($\xi<\xi_\epsilon$)
$\xi_\epsilon$ is the relevant characteristic length making the difference
with the $\epsilon=0$ case. In this region the static susceptibility, which
characterizes the linear response of the system, is given by
\begin{equation}
\chi(k,\epsilon)=1/(\Gamma_{\text{eff}} k^2+\epsilon).
\label{eq:b12}
\end{equation}
where $\Gamma_{\text{eff}}$ includes loop corrections to $\Gamma$. At the
critical state $\epsilon\rightarrow0$ we have $\chi(k,0)\sim k^{-2}$ as in
the constant force case. On the other hand, for $k\rightarrow0$ we obtain
$\chi=1/\epsilon$ so that $\chi\sim\epsilon^{-\gamma_\epsilon}$ with
\begin{equation}
\gamma_\epsilon=1.
\label{eq:b13}
\end{equation}
Eq. (\ref{eq:b12}) can be thus written as
\begin{equation}
\chi(k,\epsilon)\sim k^{-2}f(k\xi_\epsilon),
\label{eq:14}
\end{equation}
where $\xi_\epsilon\sim\epsilon^{-\nu_\epsilon}$. For $k\xi_\epsilon$ large
we should obtain $\chi(k,\epsilon)\sim k^{-2}$ so that $f(x)\sim1$ for large
$x$. On the other hand, for $k\xi$ small we should now obtain
$\chi(k,\epsilon)\sim\epsilon^{-\gamma_\epsilon}$ and therefore $f(x)\sim
x^{\gamma_\epsilon/\nu_\epsilon}$ for $x$ small with
\begin{equation}
\frac{\gamma_\epsilon}{\nu_\epsilon}=2.
\label{eq:b15}
\end{equation}
From eqs. (\ref{eq:b13}) and (\ref{eq:b15}) we thus obtain
\begin{equation}
\nu_\epsilon=1/2.
\label{eq:b16}
\end{equation}
The exponents $\gamma_\epsilon$ and $\nu_\epsilon$ results different to the
exponents $\gamma$ and $\nu$ but their ratio is the same, as one can see from
eqs. (\ref{eq:2a}) and (\ref{eq:b15}). This will carry as a consequence that
some exponents, the avalanche exponents for instance, measured in the region
$v<v_c$ can be extrapolated to the region $v>v_c$.

Finally, in the case of a local restoring force the average density of active
sites is given by
\begin{equation}
\langle\rho_a(r,t)\rangle=\frac{c}{\epsilon}.
\label{eq:b5aa}
\end{equation}
Again this result is identical to the one obtained by Vespignani {\em et al}
\cite{vespignani2} using a field theory for sandpile models. It reflects the 
balance between the driving and the local restoring forces (between driving
and local dissipation in sandpile models). Moreover this flat profile has
been observed by Barrat {\em et al} \cite{barrat} in numerical simulations of
a sandpile model with local dissipation.

The constant rate LID model thus describe a wide variety of SOC systems, like
the Barkhausen effect and sandpile models. It reproduces previous field
theory predictions for sandpile models, which we now know are also valid for
other models. Hence, the constant rate LID model provides a unifying point of
view of SOC phenomena.

\subsection{Avalanche exponents}
\label{sec:avalanches}

In the preceding section we have shown that LID models driven either at
constant force or rate belong to the same universality class. In particular
they share the same roughness and dynamic scaling exponents, which can be
estimated using previous RG calculations for the constant force LID model.
Now we proceed to show how other scaling exponents can be obtained using
$\zeta$ and $z$. For instance, we are going to compute the avalanche
exponents, which are often measured in numerical simulations of sandpile
models.

Let $s$ be the avalanche size and $T$ its duration, which are distributed
according to $P(s)$ and $P(T)$, respectively. Just at the critical state one
expect that these distributions satisfy the power law behavior $P(s)\sim
s^{-\tau_s}$ and $P(T)\sim T^{-\tau_t}$, where $\tau_s$ and $\tau_t$ are the
avalanche distribution exponents, reflecting the unexistence of
characteristic values for size and duration of the avalanches. However, for
finite $\epsilon$ and $v$ characteristic cutoffs of avalanche size $s_c$ and
duration $T_c$ will appear. These cutoffs will scale with the correlation
length (either $\xi$ or $\xi_\epsilon$ depending on the model and on the
region of the phase diagram) as $s_c\sim \xi^D$ and $T_c\sim\xi^z$, where $D$
is the avalanche fractal dimension. Near the critical state the distributions
of avalanche size and duration will thus satisfy the scaling laws
\begin{equation}
P(s)\sim s^{-\tau_s}f_1(s/s_c),\ \ \ \ P(T)\sim T^{-\tau_t}g_1(T/T_c),
\label{eq:d0}
\end{equation}
where $f(x)$ and $g(x)$ are some cutoff functions with the asymptotic
behaviors $f_1(x),g_1(x)\sim 1$ for $x\ll1$ and $f_1(x),g_1(x)\ll1$ for
$x\gg1$.

The exponents $\tau_s$, $\tau_t$, D and $z$ are not all independent. Since
$s\sim T^{z/D}$ then the condition $\int dsP(s)=\int dTP(T)$ implies
\begin{equation}
(\tau_s-1)D=(\tau_t-1)z. 
\label{eq:d1}
\end{equation}
Another scaling relation can be obtained taking into account that the
short-wavelength static susceptibility scales as the mean avalanche size
\cite{vespignani1}. From eq. (\ref{eq:d0}) we obtain $\langle s\rangle\sim 
\xi^{(2-\tau_s)D}$ while $\chi\sim\xi^{\gamma/\nu}\sim\xi^2$. Thus, setting 
$\chi\sim\langle s\rangle$ we obtain
\begin{equation}
(2-\tau_s)D=2.
\label{eq:d2}
\end{equation}
Then from eqs. (\ref{eq:d1}) and (\ref{eq:d2}) it results that
\begin{equation}
\tau_s=2-\frac{2}{D},\ \ \ \ 
\tau_t=1+\frac{D-2}{z}.
\label{eq:d3}
\end{equation}
Finally there is a scaling relation which relates the avalanche dimension
exponent $D$ with the roughness exponent $\zeta$. Below the upper critical
dimension the avalanches are compact objects \cite{paczuski2} and therefore
$s\sim \Delta h r^d$, where $\Delta h$ is the characteristic fluctuation of
the interface during the avalanche and $r$ its characteristic linear extent
in the $d$-dimensional substrate. Then, since $\Delta h\sim r^\zeta$ and
$s\sim r^D$ one obtains \cite{paczuski2}
\begin{equation}
D=d+\zeta
\label{eq:d4}
\end{equation}
Above the upper critical dimension the avalanches are no more compact and
$D=d_c=4$ \cite{paczuski2}, resulting the mean-field exponents $\tau_s=3/2$
and $\tau_t=2$.

These scaling relations are independent on the way the interface is driven.
They have been obtained assuming that $\chi\sim\langle s\rangle$, $s\sim
T^{D/z}$, there is a characteristic length $\xi$ and avalanches are compact
objects (in the interface dynamics representation).

\section{Discussion}
\label{sec:disc}

In this section we proceed to compare our results with experiments, numerical
simulations and previous theoretical works in the literature.

The average density of active sites in the stationary state was found to be
identical to the one obtained in the field theory by Vespignani {\em et al}
\cite{vespignani2}, which in principle was developed for sandpile models. 
Moreover, they predicted an upper critical dimension $d_c=4$ in agreement
with our result. Our analysis thus reveals that their approach is also valid
for other SOC models, for instance for LID models driven at constant rate.
The advantage of our approach is that it makes easier the use of momentum
space RG analysis to obtain the scaling behavior in the neighborhood of the
critical state. Using previous RG calculations for constant force LID models
we compute the complete set of scaling exponents, including the avalanche
exponents often measured in numerical simulations.

The upper critical dimension $d_c=4$ is also consistent with previous RG
analysis by D\'{\i}az-Guilera \cite{diaz-guilera} of the corresponding
Langevin equations for the height of sand columns $z_i$ (instead of $h_i$) in
the BTW model. Their results are also consistent with a dynamic scaling
exponent
\begin{equation}
z=\frac{d+2}{3},
\label{eq:e1}
\end{equation}
an expression which was previously suggested by Zhang \cite{zhang}. This
dynamic scaling is however in contradiction with the one obtained here (see
eq. (\ref{eq:3}). In his formulation D\'{\i}az-Guilera considered a coarse
graining-noise $\eta^\prime(x,t)$. In the fast time scale (that of the
evolution of the avalanches) they assumed a columnar noise uncorrelated in
space, i.e.
\begin{equation}
\langle\eta^\prime(x,t)\eta^\prime(x^\prime,t^\prime)\sim 
\delta^d(x-x^\prime).
\label{eq:e2}
\end{equation}
On the contrary, in our approach $\eta(x,h)$ is a quenched noise, a new
random variable is selected only when the interface at site $x$ advances,
i.e. when the site $x$ is active. Our choice is more appropriate because it
reflects the fact that the noise field is frozen in regions where there is no
dynamical evolution, similar to the multiplicative noise assumed in
absorbing-state phase transitions \cite{dickman}.

The estimate of the avalanche scaling exponents of sandpile models has been a
very difficult task. In two dimensions we count with a real space RG approach
by Pietronero {\em et al} \cite{pietronero}, which predicts that undirected
Abelian sandpile models, either deterministic or stochastic, belong to the
same universality class. They also provided an estimate of the avalanche size
exponent $\tau_s=1.253$. On the other hand, Priezzhev {\em et al}
\cite{priezzhev} have obtained $\tau_s=6/5=1.2$ in two dimensions, analyzing 
the structure of avalanches in the undirected Abelian sandpile model. Since
the BTW and Manna model are both Abelian \cite{dhar} this result will be
valid for both models, which are thus expected to belong to the same
universality class. Moreover, this exponent is identical to the one obtained
by De Menech {\em et al} \cite{stella} for the BTW model, using multifractal
rather than finite size scaling. These estimates can be compared with the one
obtained within our approach. In two dimensions we obtain
$\tau_s=\frac{5}{4}=1.25$ (see table \ref{tab:1}). This value is in very good
agreement with the theoretical estimates reported above. The advantage of our
approach is that we go beyond two dimensions and compute the scaling
exponents in three dimensions. Moreover we have determined the upper critical
dimension, $d_c=4$ above which MF exponents are correct.

Numerical simulations of undirected sandpile models are generally performed
with dissipation only at the boundary and assuming time scale separation. In
this case the system self-organizes into a critical state where the system
size is the only characteristic length, i.e. $\xi\sim L$. Moreover, it is
observed that the susceptibility scales as $\chi\sim L^2$. This scale
dependency has been demonstrated for Abelian models in one dimension
\cite{dhar} using the Abelian symmetry of the operators algebra. It is 
expected to be valid in any dimension as a consequence of conservation
\cite{vespignani1,vespignani2}, understanding conservation as the balance 
between the input and output of grains from the system. This scaling
dependence has been obtained here using the mapping of sandpile models to the
constant rate LID model, and it is a consequence of the balance between the
driving field and the restoring force.

The scaling exponents obtained from numerical simulations of constant force
LID models, the Manna model (the prototype of stochastic sandpile model) and
Barkhausen effect in soft-magnetic materials are shown in table \ref{tab:1}.
Some experimental estimates for the Barkhausen effect measure in soft
magnetic materials, where dipolar interactions can be neglected
\cite{durin2}, are also displayed. Our estimates using eqs. (\ref{eq:d3}) 
and (\ref{eq:d4}) are also shown for comparison. The agreement is very good,
specially in three dimensions where the RG calculations are expected to be
more precise. Moreover, we can also observe that exponents obtained for the
three different SOC models in two dimensions are very close, suggesting that
they all belong to the same universality class. In this analysis we have not
included numerical estimates for the avalanche exponents of the BTW model.
Numerical simulations of this model\cite{lubeck,chessa} lead to contradictory
results. This discrepancy may be associated with the existence of
multifractal rather than finite size scaling \cite{stella}.

This work does not cover the whole classes of SOC models. We have excluded,
for instance, directed Abelian sandpile models, critical slope models
\cite{vazquez4} and non-linear interface depinning models \cite{kardar}. 
In the case of directed Abelian sandpile models the critical exponents have
been obtained using the Abelian symmetry of the algebra of operators
\cite{dhar}. In the case of critical slope models, inspired on the work by 
Paczuski and Boettcher \cite{paczuski1}, one can perform a similar mapping
into a LID model \cite{vazquez4}. Finally non-linear interface depinning
models belongs to a different universality class described by the KPZ
equation \cite{kardar}.

\section{Summary and conclusions}
\label{sec:summ}

We have shown that undirected Abelian sandpile models and the domain wall
motion in magnetic materials where dipolar interactions are negligible can
all be mapped into the constant rate LID model and therefore display the same
universal behavior. The constant rate LID model was found to exhibit the same
critical behavior as constant force LID models above the depinning
transition. The scaling exponents were thus obtained using previous RG
analysis for constant force LID models.

In this way we have shown the existence of universal behavior in a vast class
of SOC models. We have proposed the LID driven at constant rate as the
prototype of this universality class, because it allows the use of continuous
approaches which are more treatable by analytical tools than the discrete
analysis required in most cellular automaton SOC models.

\section*{Acknowledgments}

This work has been partly supported by the {\em Alma Mater }
prize, from the Havana University.

\vskip 1in

\begin{table}
\begin{tabular}{lllllll}
$d$ & Model & $\tau_s$ & $\tau_t$ & $z$ & $D$ & Ref.\\ \hline
1 & LID & & & 1.42(3) & 2.25(1) & \cite{leschhorn}\\
  & RG & 1 & 1 & $\frac{4}{3}\approx1.33$  & 2 \\
\\
2 & LID & 1.29(2) & & 1.58(4) & 2.75(2) & \cite{leschhorn}\\
  & Manna & 1.273 & & 1.500 & 2.750 & \cite{lubeck}\\
  & Manna & 1.27(1) & 1.50(1) & 1.50(2) & 2.73(2) & \cite{chessa3}\\
  & BHN & 1.26(4) & 1.40(5) & & & \cite{durin2}\\
  & BHE & 1.28(2) & 1.5(1) & & & \cite{durin2}\\
  & RG & $\frac{5}{4}=1.25$ & $\frac{10}{7}\approx1.43$ & 
$\frac{14}{9}\approx1.56$ & 
  $\frac{8}{3}\approx2.67$ \\
\\
3 & LID & & & & 3.34(1) & \cite{nattermann}\\
  & Manna & 1.40 & & 1.75 & 3.33 & \cite{lubeck}\\
  & RG & $\frac{7}{5}=1.4$ & $\frac{7}{4}=1.75$ & $\frac{16}{9}\approx1.78$ 
& $\frac{10}{3}\approx3.33$
\end{tabular}
\caption{Scaling exponents for constant force LID models (LID), the Manna 
$d$-state model (Manna), and those obtained from experiments (BHE) and
numerical simulations (BHN) of the Barkhausen effect. Results obtained here
using RG estimates are shown for comparison.}
\label{tab:1}
\end{table}

\end{multicols}

\end{document}